# Monostable Superantiwettability

Yanshen Li[1,2], Cunjing Lv[1,3], David Quéré[4], Quanshui Zheng[1,2,5*]

[1]Department of Engineering Mechanics, Tsinghua University, Beijing 100084, China.

[2]Center for Nano and Micro Mechanics, Tsinghua University, Beijing 100084, China.

[3]Institute for Nano- and Microfluidics, Center of Smart Interfaces, Technische Universität Darmstadt, Alarich-Weiss-Straße 10, 64287 Darmstadt, Germany.

[4]Physique et Mécanique des Milieux Hétérogènes, UMR 7636 du CNRS, ESPCI, 75005 Paris, France.

[5]Applied Mechanics Lab, and State Key Laboratory of Tribology, Tsinghua University, Beijing 100084, China.

*E-mail: zhengqs@tsinghua.edu.cn

**Superantiwettability, including superhydrophobicity, is an enhanced effect of surface ruggedness via the Cassie-Baxter wetting state[1], and has many applications such as antifouling[2,3], drop manipulation[4,5], and self-cleaning[6-9]. However, superantiwettability is easily broken due to Cassie-Baxter to Wenzel wetting state transition caused by various environmental disturbances[10-17]. Since all observed reverse transitions required energy inputs[18-20], it was believed that the Cassie-Baxter state couldn't be monostable[21]. Here we show that there is a regime in the phase space of the receding contact angle and ruggedness parameters in which a Wenzel state can automatically transit into the Cassie-Baxter one without an external energy input, namely the Cassie-Baxter state in this regime is monostable. We further find a simple criterion that predicts very well experimentally observed Wenzel to Cassie-Baxter transitions for different liquids placed on various pillar-structured substrates. These results can be used as a guide for designing and engineering durable superantiwetting surfaces.**

To find rugged substrates that may have monostable Cassie-Baxter states, we use various periodical pillar-structures fabricated from a flat silicon wafer. Figure 1a shows a typical pillar-structured substrate with the pillar side length $a = 20\mu$m, height $h = 100\mu$m and separation $b = 100\mu$m. This rugged surface is treated with a commercial coating (Glaco Mirror Coat "Zero", Soft 99, Japan)[22,23] that contains hydrophobic nanoparticles (see Supplementary Section 1). The receding and advancing contact angles of water on a flat silicon surface with this coating are measured as 150 ± 3° and 164 ± 2°, respectively. Figure 1b shows a side view of a water drop placed on this substrate, appearing clearly in the Cassie-Baxter state (namely the liquid rests on the tops of the pillars with trapped air under the liquid). We then press the drop on its top by a copper plate coated with superhydrophobic CuO nanostuctures to transit the wetting state of the drop into a Wenzel state (see Fig. 1c, namely the liquid has intimately contacts with the substrate surface, and see the experimental setup in Supplementary Section 2 ). Starting from this state, we lift the plate at a very low and constant speed (20 μm/s). Figures 1d-1i show six sideviews captured from a recorded movie (Supplementary Video 1) of the drop-substrate-plate system during a plate-lifting process. With increasing the distance, $z$, between the plate's bottom and the pillars' tops, we observe that the contact of the drop on the substrate experiences successively three stages. In Stage 1, the contact radius, $R_s$, of the drop is decreasing while the whole contact area is Wenzel's (Fig. 1d). In Stage 2, the $R_s$ is locked but the contact area initiates transiting into Cassie-Baxter's at the contact edge (Fig. 1e), then expends the transited area (Fig. 1f-1g) until the contact suddenly becomes the fully Cassie-Baxter's (Fig. 1h). We name this stage the Wenzel to Cassie-Baxter (W2C) transition stage. In Stage 3, with further lifting the plate the drop keeps Cassie-Baxter contact and the $R$ is continuously reducing (Fig. 1i) until the drop takes off from the plate. The final Cassie-Baxter state is the same as the initial state (see Supplementary Section 3). Figure 2a shows the dependence of the contact radius $R_s$ upon the distance $z$, in which the two vertical dashed lines separate the three stages.

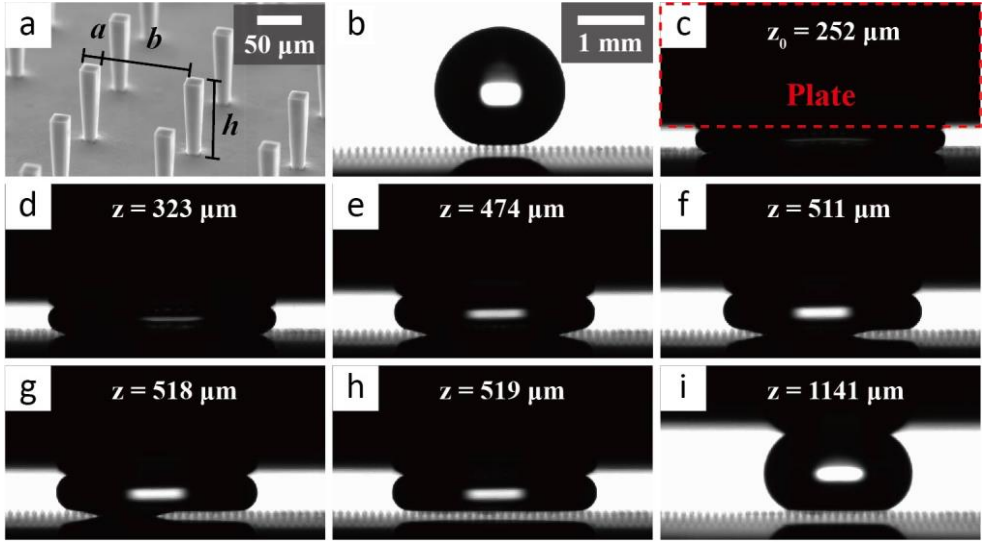

**Figure 1 | The Wenzel to Cassie-Baxter (W2C) transition process**. **a**, The oblique view of a typical pillar-structured substrate with the pillars' height $h = 100$ μm, side length $a = 20$ μm, and separation distance $b = 100$ μm, fabricated from a silicon wafer and then coated with the commercial agent Glaco. **b**, A water drop of diameter 1.96mm (smaller than its capillary length 2.71 mm so that the gravidity effect is negligible) placed on this substrate, appearing clearly in the Cassie-Baxter state. **c**, The drop is then forced in a Wenzel state by pressing the drop using a superhydrophobic (nano-structured CuO) plate. Starting from this state, the plate is being gradually lifted at at a constant very slow speed ($v = 20$ μm/s), where $z = vt$ denotes the lifted distance. **d-i**, Selected succesive side views of the drop-substrate-plate system captured from a SI movie in a lifting process. **d**, The W2C transition initiates from the edge of the drop-substrate interface. **e-g**, The W2C transitted zone gradually extends to the central interface area until **h**, it suddenly extends to the whole interface. **i**, Further lifting the plate doesn't affects the Cassie-Baxter mode.

During the lifting, energy input could be needed to overcome possible energy barrier against the W2C transition. The inevitable contact angle hysteresis will result in energy dissipation in moving the three-phase (liquid-solid-air) contact lines on both the substrate and the plate and thus requires more energy to realize the W2C transition. However, if we can prove that the resultant pressing force *F* exerted by the drop on the plate could be always positive during the W2C transition process, then the drop-substrate-plate system continuously outputs energy. Thus the Cassie-Baxter state must be monostable.

With the above idea in mind, we measure the force *F*. In our experiments, the *F* can be easily estimated through the precise relationship $F = \pi R_1^2 \gamma (R_2^{-1} - R_1^{-1})$ (see derivation in Supplementary Section 4) by measuring the two principal curvature radii, $R_1$ and $R_2$, of the drop at the drop's great circle, as illustrated in the Insert of Fig. 2b. At the same time the Laplace pressure - the drop's internal pressure induced by the drop's surface tension can be calculated as $p = \gamma(R_1^{-1} + R_2^{-1})$. We plot in Figs. 2b and 2c the respective dependences of the measured $F$ and $p$ upon the distance $z$. It is observed that $p$ as well as $F$ are nearly constant in the whole W2C transition stage. Furthermore, during the whole W2C transition stage the force $F$ is indeed positive, indicating that the drop is continuously outputting mechanical work $W(z) = \int_{z_0}^{z} F dz$ to the plate. Consequently, the excess free energy of the drop $U(z) = -W(z)$ is monotonously decreasing with $z$ (Fig. 2d). Here, we ignore the internal fluidic dissipation of the drop during the quasi-static moving (with respect to the very slow lifting speed) and the gravity potential change (for the very small drop, diameter 1.96 mm). Therefore, the above result proves that the studied Cassie-Baxter state is monostable. As further increasing $z$ beyond the end of W2C transition stage, we observe from Fig. 2d the continuously decreasing of the free energy until its minimum (at $z$ = 2913 μm). After that, the free energy increases a little until the drop detaches the plate.

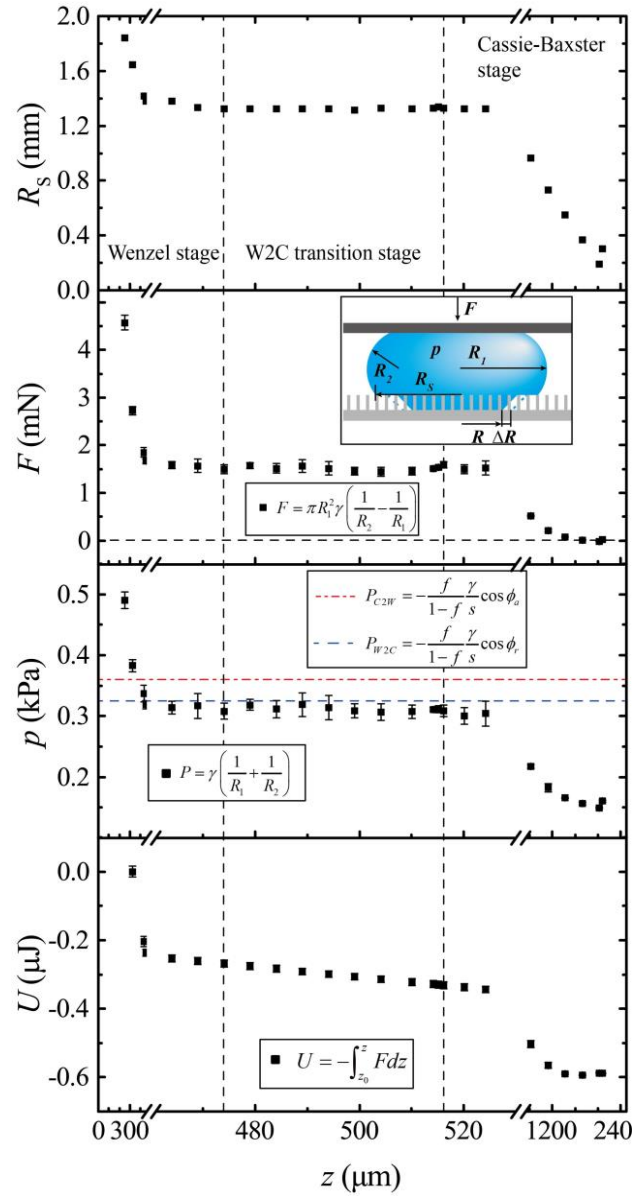

**Figure 2 | Press force and free energy of the drop**. a. contact radius of the drop and the surface of the pillars top, $R_s$, decreases as lifting the plate. At some point, $R_s$ do not decrease before the drop is in Cassie state, we denote this point z = 474 μm as the start of the transition process. Before this, it is called the Wenzel stage. Further lifting the plate, the drop turn into Cassie state at z = 516 μm, we call this period the W2C transition stage. The drop stays in Cassie state thereafter, which is called the Cassie stage. **b**, The press force, $F$, can be calculated through measrung the two principal curvature radii $R_1$ and $R_2$ at the drop's great circle (see the Insert). The dots show the measured dependence of $F$ upon the lifting distance z. The left vertical dotted line marks where we observe the start of the W2C state transition, and the right one marks the end of the W2C state transition. c. dependance of Laplace pressure within the drop on $z$. The W2C pressure is measured and found to agree well with equation (2). **d**, The dots show the free energy of the drop as a function of $z$, that confirms the monostability of the free Cassie-Baxter state. Error bars of $F$, $p$, $U$ indicate the standard deviation of five measurements.

With the W2C transited zone expanding (or equivalently, the radius of the none transited zone reducing from $R$ to $R - \Delta R$), the free energy changes (see Supplementary Section 5 for detailed derivation)

$$\Delta U = \left[(1-f)+(r-f)\cos\phi_r\right]\gamma(2\pi R\Delta R)$$

where $f$ and $r$ are the areas of the pillars' tops and the rough substrate per unit apparent substrate area, respectively, and $\phi_r$ denotes the receding contact angle of the substrate material. Since the drop is receding, the receding contact angle, $\phi_r$, rather than the Young's contact angle is involved, and the energy dissipation in moving the three-phase contact lines has been accounted through the use of $\phi_r$. A sufficient and necessary condition for realizing automatically W2C transition is $\Delta U<0$ during the whole transition process. Therefore, from the above relation we obtain the following criterion of monostable Cassie-Baxter state:

$$\frac{1-f}{r-f} < -\cos\phi_r \tag{1}$$

Interestingly, the criterion for a globally stable Cassie-Baxter state (i.e. it has a lower free energy than that in Wenzel's) has a very similar form $\frac{1-f}{r-f} < -\cos\phi_Y$ to Eq. (1)[24-26]. Because the Young's contact angle $\phi_Y$ is always larger than the receding angle $\phi_r$, the globally stability is, not surprisingly, a necessary condition for realizing the monostability.

To validate the criterion (1), we conduct a series of similar experiments to that as shown in Fig. 1 with varying $\phi_r$, $f$, and $r$. In the experiments, silicon substrates with square or circular pillars of fixed side length or diameter $a=100$ μm and height $h=100$ μm are used, and the space between pillars $b$ is varied to change $f = a^2/(a+b)^2$ or $f = \pi a^2/4(a+b)^2$ (see Supplementary Section 6). To have a wider choice of $\phi_r$, we use the mercury, instead of water, as the testing liquid. Different

**7 / 17**

surface treatments to the substrates are used to change the contact angles of the mercury drops (See Methods section). The open and solid dots in square (for square pillars) and circular (for circular pillars) shape in Fig. 3a show the results that are experimentally observed to be monostable and non-monostable, respectively. In comparison, the green area above the dashed line in Fig. 3a corresponds to the criterion (1). As can be seen, the criterion (1) has an excellent agreement with the experimental observations.

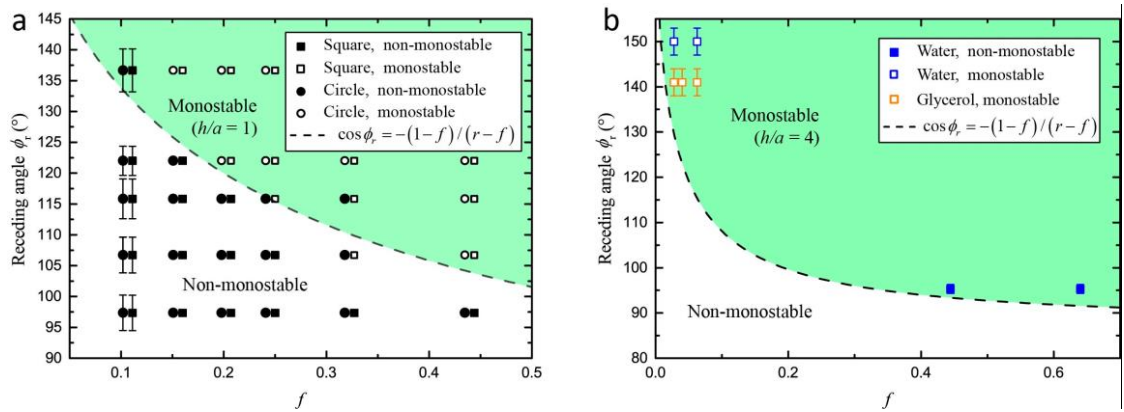

**Figure 3** | Experimentally observed monostable Cassie-Baxter (open dots) and metastable Cassie-Baxter (solid dots) cases in the $(\phi_r, f)$ phase space on pillar-structured substrates fabricated from silicon wafer. (a), The pillars are square-shaped with fixed height $h=100$ μm and side length $a=100$μm, the pillars' separation $b$ is varied to give different area fraction $f = a^2/(a+b)^2$ and roughness $r = 1 + \frac{h}{s}f$. Each row consists of six data points, corresponding to six small substrates. They are taken from one big ion etched silicon wafer, and are treated at the same time, so their surface share the same property (including contact angle). Receding contact angles are averaged over six measurements, and the error bars indicate standard deviations of the data. (b), experiements for water and glycerol. The pillars are square-shaped with fixed height $h=100$ μm, side length $a=100$ μm, and different separations $b$. Receding contact angles are averaged over five measurements. Error bars indicate standard deviations of the data.

The criterion (1) should also be generally valid for various liquids and rugged substrates. To find supporting evidences, we adopt the fixed $a = 20\mu$m, $h = 100\mu$m, and varying $b$ for the pillar-structured substrates so that the area frictions $f$ are similar to those used in Fig. 3a but the ruggedness $r = 1 + \frac{h}{s}f$, or equivalently the height/side length ratio $\frac{h}{a}$ are different. Figure 3b shows the experimental results with open or solid dots indicating monostable or non-monostable

Cassie-Baxter states for water (blue dots) and glycerol (orange dots, see Supplementary Section 7 for ). The green area above the dashed line in Fig. 3b is the monstable Cassie-Baxter state zone predicted by the criterion (1). Again, we see an excellent agreement between the theoretical and experimental results. Unlike mercury, for water and glycerol we are challenged to have more contact angles for the experiments.

To understand why the observed Laplace pressures, $p_{W2C}$, are constant during the whole W2C transition stage, we first note that there is another critical Laplace pressure, $p_{C2W}$, the maximum pressure before the pillars will pierce into the liquid (or equivalently the Cassie-Baxter state will transit into the Wenzel state), that can be precisely formulated as [27]

$$p_{C2W} = -\frac{f}{1-f}\frac{\gamma}{s}\cos\phi_a \qquad (2)$$

where $\phi_a$ denotes the advancing contact angle of the substrate's surface, and $s$ denotes the ratio of the area to perimeter of the pillars' cross-sections. This result is valid for any cross-section shaped pillar-structures. In particular, for both square and circular-shaped cross-sections with side length or diameter $a$, it is easy to get $s = a/4$. The observation $p_{W2C} < p_{C2W}$ as indicated by the red dash-dotted line in Fig. 2c is consistent with the physical concepts of $p_{W2C}$ and $p_{C2W}$. After a small expending of the W2C transited zone, a little amount of liquid will be drained away from under the pillars' top to the above drop, and thus increase the Laplace pressure until the plate will be further lifted up. In this interval, the increased Laplace pressure will stop the W2C expanding but is still lower than the critical pressure $p_{C2W}$ so that the W2C transited zone can persist. After a certain further moving up of the plate, the Laplace pressure will be lowered down again to reach the value $p_{W2C}$, and consequently leads to the W2C transition expansion again. We therefore call $p_{W2C}$ the W2C transition pressure. Since in the W2C transiting zone the process of the liquid surface is inverse to that in the piercing one, and is thus a receding motion, using a similar derivation (see Ref. 27, we can obtain

$$P_{W2C} = -\frac{f}{1-f}\frac{\gamma}{s}\cos\phi_r. \qquad (3)$$

This model prediction as indicated by the blue dashed line in Fig. 2c agrees excellently well with the experimental results.

All the above-reported results don't consider the effect of drop's size. In fact, we experimentally find that the $p_{W2C}$ is independent of liquid drop volume in most cases. However, for a very small drop with radius $R_{sph}$ as the drop of the same volume is spherical, the Laplace pressure $p_{sph} = \frac{2\gamma}{R_{sph}}$ can exceed the W2C transition pressure (3). Since in the Cassie-Baxter state the drop is nearly spherical, we therefore obtain the following critical radius that drops much have larger radii in order to realize a monostable Cassie-Baxter state:

$$R_{sph,cr} = -\frac{1-f}{f}\frac{a}{2\cos\phi_r} \qquad (4)$$

This property is independent of liquid and pillars' height. For example, as $f = 0.1$ and $\phi_r = 130°$ or $150°$, we obtain from (4) that $R_{sph,cr} = 7.0a$ or $5.2a$, respectively. Thus, smaller $a$ and $f$ and larger $\phi_r$ allow smaller drops in monostable Cassie-Baxter contact.

The criterion (1) points out that we can use a higher area friction $f$ for lowering down the requirement of a large $\phi_r$ in order to achieve a monostable superantiwetting substrate. However, this will reduce the apparent contact angle, $\phi^*$, on the rugged surface according to the Cassie-Baxter relationship: $\cos\phi^* = -1 + (1+\cos\phi_Y)f$; while a super-large contact angle is needed for achieving superantiwettability. The criterion (1) also indicates that we can use a large ruggedness $r$ (or equivalently larger pillars' height to side length ratio because of $r = 1 + \frac{h}{s}f$) $h/a$ to lower down the requirement for $\phi_r$. This will however increase the fabrication cost of the pillar-structure, and enlarge the rise of mechanical instability of the slender pillars.

Therefore, for achieving an optimized pillar-structure, we should find an additional tunable parameter. The criterion (1) contains only two dimensional ruggedness parameters: $f$ and $r$, and is thus size-independent. We note that wetting along the edge of the foot of a pillar must be different from wetting on the pillar's top and side. Involving this effect, following a similar approach as suggested by Quanshui Zheng[28], we can obtain the following refined criterion instead of (1):

$$\frac{1-(1+\lambda_{edge}/s)f}{r-f} < -\cos\phi_r, \tag{5}$$

where $\lambda_{edge} = \sigma_{edge}/\gamma$ is a length-dimensional material parameter with $\sigma_{edge}$ denoting the line energy along the edge of the pillar food. For water and pillar-structured coated with octadecyltrichlorosilane (OTS, $C_{18}H_{37}Cl_3Si$, 95%), $\lambda_{edge}$ is estimated in the submicron range. Nevertheless, from (4) we see that through proportionally shrinking all the pillar-structural parameters $a$, $b$, and $h$, both $f$ and $r$ are invariant, but $a$ (and $s$) is shrinking. This approach yields lowering down the requirement to $\phi_r$ according to (5), and simultaneously increasing the critical pressure $p_{cr}$, enlarging the apparent contact angle through the refined Cassie-Baxter relationship[28]: $\cos\phi^* = -1 + (1+\cos\phi_Y - 4\lambda_{edge}/a)f$, and boarding the valid range of the drops' sizes as shown in (3). The latter feature is particularly useful when consider applications in condensation and antifogging. In these cases drops are initiated from nano-sized.

In summary, the above-reported results can help to explore, design, and optimize monostable superantiwetting substrates for various liquids. Water is among the most interested. Unfortunately, the Young's contact angles of water rarely exceed 120 °[29], that is very close to the lower bound of contact angle in the phase diagram (Fig. 3); while the receding angles of water are even lower (<100 °). Nevertheless, the refined criterion (5) may guide a way in a near future to find hierarchical rugged structures that are monostably superantiwetting for water drops formed even though in condensation.

# Methods

## Fabrication of superhydrophobic two-tier roughness surfaces.

Silicon substrates with square pillars are fabricated by photolithography and deep reactive-ion etching (DRIE). Then the silicon substrates are treated by a commercial coating (Glaco Mirror Coat "Zero", Soft 99, Japan) composed of hydrophobic nanoparticles and volatile liquid. Substrates are dipped into the solution and then dried in air. After that, the substrates are put into an oven and kept at 150°C. The dipping and heating processes is repeated three to four times.

## Fabrication of superhydrophobic CuO nanostructured surface

Fabrication of superhydrophobic CuO nanostructure on copper bricks is composed of three steps: cleaning, oxidation and silanization, as described by Nenad[30].

Cleaning: The copper brick was first ultrasonically cleaned in acetone for 10 min, and then in ethanol, isopropanol, and deionized (DI) water successively. The brick was then dipped into a 2.0 M hydrochloric acid solution for 10 min to remove the native oxide film on the surface, then triple-rinsed with DI water, and dried with clean nitrogen gas.

Oxidation: nanostructured CuO films were formed by immersing the cleaned bricks into a hot ($96 \pm 3°C$) alkaline solution composed of $NaClO_2$, $NaOH$, $Na_3PO_4 \cdot 12H_2O$, and DI water (3.75:5:10:100wt%) for 20 min. During the oxidation process, a thin (≈300 nm) $Cu_2O$ layer was formed that then reoxidized to form sharp, knife-like CuO oxide structures, see Figure 4a. Then the bricks were triple rinsed in DI water and dried with clean nitrogen gas.

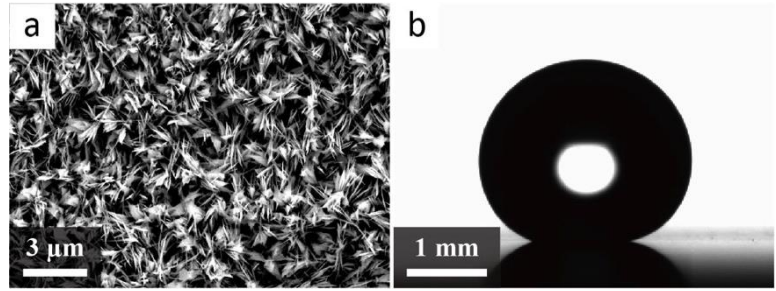

Figure 4. (a): SEM image of the CuO nanostructure, and (b): the contact angle of a drop on this silanized CuO nanostructured surface.

Silanization: After oxidation, the nanostructured CuO was oxygen plasma (Femto SLCE, Diener Electronic, Germany) cleaned at 30 W for 2 hours. Then the bricks were immediately placed in a solution composed of Hexadecane and Trichloro(octadecyl)silane (25:0.1 volume) for 20 min. The samples were then dipped in chloroform for 15 min to remove any residual organics or liquids. Then they were dipped in ethanol for 15 min and dried with clean nitrogen gas.

After silanization, the CuO nanostructure has a typical water receding angle of $\phi_r = 145 \pm 3°$, as shown in Figure 4b.

For experiments with mercury, the step of silanization is not necessary.

## Treatment of the silicon substrate to change its contact angle with mercury

In order to change the contact angle of mercury on silicon, several methods of surface treatment are used. The methods and the corresponding contact angles are listed in table 1.

Table 1. Treatments and the corresponding contact angles of mercury

| No. | Method | Advancing CA | Receding CA |
|---|---|---|---|
| 1 | OTS | 165 ± 2° | 137 ± 4° |
| 2 | None | 144 ± 4° | 123 ± 3° |
| 3 | Plasma 1 | 144 ± 1° | 116 ± 3° |
| 4 | Plasma 2 | 137 ± 2° | 107 ± 3° |
| 5 | Plasma 3 | 132 ± 1° | 97 ± 3° |

**OTS**: the Si micro-structured surface is ultrasonically cleaned in acetone for 10 min, and then in ethanol and deionized (DI) water successively. The substrate is dried by clean nitrogen. Then it is silanized by the same procedure as for CuO.

Plasma 1: after treated by the method "Plasma 2", the sample is put in clean air at room temperature for one week.

Plasma 2: after cleaning, the substrate is plasma cleaned at 30 W for 30 min. Where in this case, air is used.

Plasma 3: after cleaning, the substrate is oxygen plasma cleaned at 30 W for 30 min. Where in this case, pure oxygen is used.

## Acknowledgements:

Financial support from the NSFC under grant No. 11372153 and 11172149 and financial support from Statoil ASA (Norway) through the project nanotechnology for anti-icing application is gratefully acknowledged. The authors thank David Quéré for fruitful suggestions and Shuai Wu for discussions, Dong Huang and Pengfei Hao for providing some of the samples.

## Author contributions:

Q.S.Z. and Y.S.L. proposed and designed the project. Y.S.L. conducted the experiments and data analysis. Q.S.Z. and Y.S.L performed theoretical analysis. Y.S.L. and Q.S.Z. wrote the manuscript, C.J.L. participated in manuscript preparation and theoretical analysis.

## Competing financial interests

The authors declare no competing financial interests.